\documentclass[11pt]{article}
\usepackage{latexsym}
\usepackage[dvips]{graphics}

\newcommand{\be}{\begin{equation}}
\newcommand{\ee}{\end{equation}}
\newcommand{\bea}{\begin{eqnarray}}
\newcommand{\eea}{\end{eqnarray}}
\newcommand{\Tr}{\mbox{Tr}}

\newcommand{\bc}{\begin{center}}
\newcommand{\ec}{\end{center}}

\newcommand{\prd}{Phys.\ Rev.\ D }
\newcommand{\ov}{${\cal O}(v^6)$ }

\begin{document}

\title{Unquenched Charmonium with NRQCD}
\author{Chris Stewart and Roman Koniuk \\
Department of Physics \& Astronomy \\ York University, 4700 Keele
St., Toronto, Ontario M3J 1P3, Canada.}
\date{\today}

\maketitle


\begin{abstract}
We present the results from a series of lattice simulations of the
charmonium system using a highly-improved NRQCD action, both in
the quenched approximation, and with $n_{f} = 2$ light dynamical
quarks. The spectra show some evidence for quenching effects of
roughly $10 \%$ in the $S$- and $P$-hyperfine spin
splittings---probably too small to account for the severe
underestimates in these quantities seen in previous quenched
charmonium simulations. We also find estimates for the magnitude
of other systematic effects---in particular, the choice of the
tadpole factor can alter spin splittings at the $10$--$20 \%$
level, and ${\cal O}(\alpha_{s})$ radiative corrections may be as
large as $40 \%$ for charmonium. We conclude that quenching is
just one of a collection of important effects that require
attention in precision heavy-quark simulations.
\end{abstract}

\unitlength=1.0cm


\section{Charmonium on the lattice}

One of the most rapidly expanding sectors of lattice QCD in the
last decade has been the study of heavy-quark systems. Lattice
simulations have successfully reproduced the broad structure of
the heavy hadron spectrum, providing a solid piece of evidence for
the correctness of QCD. Discrepancies at the level of the
hyperfine structure still persist however, and in some cases these
are uncomfortably large.

This paper describes a series of highly-improved non-relativistic
simulations of the charmonium system, with the aim of estimating
the sizes of various systematic uncertainties influencing the
spectrum. An understanding of the relative influence of these
uncertainties on the heavy-quark spectrum is vital to the
interpretation of the current state of lattice simulations.

One very successful approach to simulating heavy quark systems
utilises the NRQCD formalism \cite{NRQCDOriginal,NRQCDImproved},
where the quark dynamics are governed by an effective
non-relativistic Hamiltonian, expanded in powers of the
heavy-quark velocity. For the bottom and charm quarks, $v^{2} \sim
0.1$ and $v^{2} \sim 0.3$ respectively, and so we expect to
achieve some success with a non-relativistic theory. Simulations
of heavy-light and heavy-heavy charm and bottom systems have shown
that NRQCD captures much of the correct physics of the heavy
quarks. Understanding the remaining systematic errors in
heavy-quark simulations has become a major focus of the lattice
NRQCD community.

The first report of a high-statistics NRQCD simulation
of charmonium appeared in 1995 by Davies \emph{et al.} \cite{DaviesCharm}. The
authors used a NRQCD Hamiltonian with relativistic and
discretisation errors corrected to ${\cal O}(v^{4})$ to measure ground and excited
$S$, $P$ and $D$ state energies in the quenched approximation. Agreement
with experiment was very promising, with discrepancies at the order of
$10$--$30\% $ in $S$- and $P$-state hyperfine spin-splittings,
in agreement with the expected size of the next-order corrections.

Disturbingly, charmonium simulations incorporating ${\cal
O}(v^{6})$ corrections \cite{TrottierQuarkonium} showed a large
\emph{decrease} in hyperfine spin splittings, taking theoretical
predictions as much as $50\%$ further away from experimental
values. These simulations also demonstrated a large dependence on
the definition of the tadpole correction factor. Given the size of
$v^{2}$ for charmonium, sizeable ${\cal O}(v^{6})$ corrections
are not surprising; however, the disappointingly large
discrepancies in the spectrum with such a highly improved theory
give pause to the future of charmonium simulations. Evidently, the
NRQCD expansion converges slowly for the charm quark.

Even in the less-relativistic $\Upsilon$ system, the same
highly-improved NRQCD action has not provided conclusive agreement
with experiment \cite{HighImpUps,UnqUps}. Certainly, NRQCD to ${\cal O}(v^{6})$
is not a closed problem.

The difficulties with the hyperfine spectrum are not limited to
the NRQCD approach. A report on the status of charmonium
simulations with the relativistic Fermilab approach in 1993
\cite{charm92} cited a $20$--$30 \%$ shortfall for the $S$-state
hyperfine splitting using an SW-improved Wilson action. In 1999,
the UKQCD collaboration reported on a tadpole- and SW-improved
simulation of charmonium \cite{UKQCDRelCharm99}; their results for
the $S$-hyperfine splitting were roughly $40 \%$ below the
experimental value. Both of these simulations used the quenched
approximation, and the inclusion of dynamical quark loops would
increase the hyperfine splittings. In the 1993 report, quenching
effects were estimated to be as large as $40\%$, however this
seems optimistic---corrections at the $5$--$15\%$ level seem more
typical in full QCD simulations of both the $\Upsilon$ system
\cite{UnqUps} and of light hadrons \cite{FermionLoops}.

A very recent report from the CP-PACS collaboration
\cite{CPPACS00} describes unquenched simulations of charmonium and
bottomonium using NRQCD over a range of lattice spacings and
sea-quark masses, with $n_{f} = 2$ SW-improved light sea-quarks.
In that report, the authors concentrate mostly on simulations of
the $b \bar{b}$-system, though some charmonium results are
presented. Their results indicate a significant increase in the
$S$-state hyperfine splittings as the sea-quark mass decreases
towards the chiral limit, though no effect is seen on the
$P$-states.

We have performed a series of highly-improved NRQCD simulations to
examine the various systematic uncertainties influencing the
charmonium spectrum. We first concentrate on the effects of
dynamical quark loops. If these account for the majority of the
hyperfine splitting discrepancy in charmonium then we expect to
find a large increase in the splittings when dynamical quarks are
included, even in the NRQCD formalism. We examine this effect
using an ensemble of unquenched configurations provided by the
MILC collaboration, seeking to establish whether the effects of
dynamical quarks are sufficient to reconcile the hyperfine
discrepancy. This work does \emph{not} aim to provide the
definitive unquenched charmonium spectrum.

The remainder of the paper is devoted to an examination of other
systematic effects. Simulations with two common definitions of the
tadpole correction factor result in significantly different
spectra, and we find a rough estimate of the effect of ${\cal
O}(\alpha_{s})$ radiative corrections to the NRQCD expansion
coefficients. Finally, we note a sizeable shift in the hyperfine
splittings due to an instability in the standard form for the
heavy-quark propagator's evolution equation. Each of these effects
is contrasted with the estimated magnitude of the unquenching
error, which leads us to several conclusions about NRQCD
simulations of charmonium, and heavy-quark simulations in general.


\section{The standard lattice NRQCD formalism}
\label{Theory}

The NRQCD Hamiltonian is typically presented as an expansion in powers
of the heavy-quark velocity. A highly-improved NRQCD
Hamiltonian, with corrections to ${\cal O}(v^{6})$ in the velocity expansion
\cite{NRQCDImproved}, is
\be
\label{NRQCDHamiltonian}
H = H_{0} + \delta H_{v^{4}} + \delta H_{v^{6}} \, ,
\ee
where
\be
H_{0} = \frac{-{\bf \Delta}^{(2)}}{2M_{0}} \, ,
\ee
is the leading kinetic Schr\"odinger operator, and the ${\cal
O}(v^{4})$ and ${\cal O}(v^{6})$ corrections are
\bea
\delta H_{v^{4}} &=& -c_{1} \frac{1}{8M^{3}_{0}}
\left({\bf \Delta}^{(2)}\right )^{2}
+ c_{2} \frac{ig}{8 M^{2}_{0}} \left ( \tilde{{\bf \Delta}} \cdot
\tilde{{\bf E}} - \tilde{{\bf E}} \cdot \tilde{{\bf \Delta}} \right ) \nonumber \\
& & + c_{3} \frac{g}{8 M^{2}_{0}} \sigma \cdot \left ( \tilde{{\bf \Delta}}
\times \tilde{{\bf E}} - \tilde{{\bf E}} \times \tilde{{\bf \Delta}} \right )
- c_{4} \frac{g}{2M_{0}} \sigma \cdot \tilde{{\bf B}} \nonumber \\
& & + c_{5} \frac{a^{2}}{24 M_{0}} {\bf \Delta}^{(4)}
- c_{6} \frac{a}{16s M_{0}^{2}} \left({\bf \Delta}^{(2)}\right )^{2} \, , \\
\delta H_{v^{6}} &=& - c_{7} \frac{g}{8M^{3}_{0}}
\left \{ \tilde{{\bf \Delta}}^{(2)}, \sigma \cdot \tilde{{\bf B}} \right \}
\nonumber \\
& & - c_{8} \frac{3g}{64M^{4}_{0}}
\left \{ \tilde{{\bf \Delta}}^{(2)}, \sigma \cdot \left ( \tilde{{\bf \Delta}}
\times \tilde{{\bf E}} - \tilde{{\bf E}} \times \tilde{{\bf \Delta}} \right ) \right \}
\nonumber \\
& & - c_{9} \frac{ig^{2}}{8M^{3}_{0}} {\bf \sigma} \cdot \tilde{{\bf E}} \times
\tilde{{\bf E}} \, .
\eea
A \emph{tilde} signifies the use of improved versions of the
lattice operators that remove the leading discretisation
errors: the improved lattice derivatives $\tilde{\Delta}$ and
$\tilde{\Delta}^{(2)}$ are given by
\bea
\label{DiscreteSecondOrder}
\tilde{\Delta}_{\mu} \psi(n) &=& \Delta_{\mu} \psi(n) - \frac{a^{2}}{6}
\Delta_{\mu}^{3} \psi(n) \, , \nonumber \\
\tilde{\Delta}^{2}_{\mu} \psi(n) &=& \Delta^{2}_{\mu} \psi(n) + \frac{a^{2}}{12}
(\Delta^{2})^{2} \psi(n) \,,
\eea
while the fields $\tilde{E}_{i} = \tilde{F}_{4i}$ and
$\tilde{B}_{i} = \frac{1}{2} \epsilon_{ijk} \tilde{F}_{jk}$ are taken
from an improved gauge field tensor \cite{TrottierQuarkonium,NRQCDImproved},
\bea
\tilde{F}_{\mu\nu}(n) &=& \frac{5}{3} F_{\mu\nu}(n)
- \frac{1}{6} \left [U_{\mu}(n) F_{\mu\nu}(n+\hat{\mu})
    U_{\mu}^{\dag}(n)
    + U_{\mu}^{\dag}(n-\hat{\mu}) F_{\mu\nu}(n-\hat{\mu})
    U_{\mu}(n-\hat{\mu}) \right . \nonumber \\
& & \left . - U_{\nu}(n) F_{\mu\nu}(n+\hat{\nu})
    U_{\nu}^{\dag}(n)
    + U_{\nu}^{\dag}(n-\hat{\nu}) F_{\mu\nu}(n-\hat{\nu})
    U_{\nu}(n-\hat{\nu}) \right ] \, .
\eea
All lattice operators are tadpole improved \cite{LepageImproved}, by
dividing all instances of the link operators $U_{\mu}(n)$ by the
tadpole correction factor $u_{0}$,
\be
U_{\mu}(n) \to \frac{U_{\mu}(n)}{u_{0}} \, .
\ee
This means, for example, that the
gauge {\bf E} and {\bf B} fields are adjusted by a factor of
$u_{0}^{4}$. Much evidence exists for the superiority of the
\emph{Landau} definition of the tadpole factor,
\be
u_{0}^{L} = \left \langle \frac{1}{3} \Tr U_{\mu} \right
\rangle_{\partial_{\mu} A_{\mu} = 0} \, ,
\ee
over the \emph{plaquette} definition,
\be
u_{0}^{P} = \left \langle \frac{1}{3} \Tr P_{\mu \nu} \right
\rangle^{1/4} \, .
\ee
For example, $u_0^L$ leads to smaller corrections to hyperfine
splittings, and better scaling of quarkonium masses
\cite{TrottierQuarkonium,Howard}; it restores rotational
invariance to a greater degree in the static quark potential
\cite{NormSU2}; and it results in closer agreement between the
tadpole-improved value and the perturbative value for the `clover'
coefficient $c_{sw}$ in the Sheikholeslami-Wohlert action
\cite{LuscherCsw}. We have used both the Landau and plaquette definition
in our simulations.

Since the quarks and antiquarks are decoupled in the non-relativistic
theory, the heavy-quark Green's function may be found from an evolution
equation,
\be
\label{EvolutionEquation}
G_{t+1} = \left ( 1-\frac{aH_{0}}{2s} \right )^{s} U_{4}^{\dag}
\left (1-\frac{aH_{0}}{2s}\right)^{s} \left (1-a \delta H \right)
G_{t} \, ,
\ee
with the initial time-step given by
\be
G_{1} = \left (1-\frac{aH_{0}}{2s}\right)^{s} U_{4}^{\dag}
\left (1-\frac{aH_{0}}{2s}\right)^{s} \delta_{x,0} \, .
\ee
The $\left ( 1-aH \right )$ factors are linear approximations to the continuum
evolution operator $e^{Ht}$. The `stabilisation parameter' $s$
appearing in Equations (\ref{NRQCDHamiltonian}) and
(\ref{EvolutionEquation}) improves the approximation to the time
evolution operator $e^{aH}$.

To complement the use of a highly-improved quark Hamiltonian, we use
a tadpole and `rectangle' improved action for the gauge fields
\cite{LepageImproved},
\be
\label{ImpGaugeAction}
S_{G} = - \beta \sum_{n,\mu>\nu} \left (
\frac{5}{3u_{0}^{4}} P_{\mu\nu}(n)
 - \frac{1}{12u_{0}^{6}} (R_{\mu\nu} + R_{\nu\mu}) \right ) \, ,
\ee
where $P_{\mu\nu}(n)$ and $R_{\mu\nu}$ represent the traces of $1 \times 1$
plaquettes and $2 \times 1$ rectangles of link operators respectively.

Operators for the various quarkonium states have the form
\be
M(t) = \sum_{n} \psi^{\dag}(n,t) \,
\Gamma(n) \, \chi^{\dag}(n,t) \, ,
\ee
where $\psi^{\dagger}$ and $\chi^{\dagger}$ are the quark and antiquark
creation operators, and $\Gamma(n)$ provides the appropriate spin and spatial
wavefunction
quantum numbers. Operators for the lowest-lying $S$, $P$ and $D$ states are
given in a number of references \cite{DaviesCharm,TrottierQuarkonium};
using these, we have constructed
propagators for each of the $^{2S+1}L_{J} = \, ^{1}S_{0}$,
$^{3}S_{1}$, $^{1}P_{1}$, $^{3}P_{0}$, $^{3}P_{1}$ and $^{3}P_{2}$ states.
Only one spin polarisation of each of the triplet states was used.

To reduce the effects of excited-state contamination and improve the
operators' overlap with the true meson ground-state wavefunctions, we have
used a gauge-invariant smearing function, replacing
\be
\Gamma(n) \to \Gamma(n) \phi_{sm}(n) \, .
\ee
A simple and effective choice for $\phi_{sm}$ is \cite{smearing}
\be
\label{smearing}
\phi_{sm}(\epsilon,n_{s}) = \left ( 1 + \epsilon \Delta^{2} \right
)^{n_{s}} \, .
\ee
The weighting factor $\epsilon$ and number of smearing iterations
$n_{s}$ were tuned to optimise the overlap with the ground state.

\section{Details of the simulations}

We have performed a number of different simulations of the charm
system, to compare the magnitudes of various systematic effects
on the spectrum. We obtained results with the NRQCD
Hamiltonian in Equation (\ref{NRQCDHamiltonian}) truncated to
${\cal O}(v^{4})$ and ${\cal O}(v^{6})$, with both the Landau and
plaquette definitions for the tadpole factor $u_{0}$.

To examine the size of dynamical quark effects, we obtained
an ensemble of 200 unquenched gauge field configurations, generously
provided by the MILC collaboration \cite{MILCConfigurations}.
The configurations were created with the
Wilson gluon action at $\beta = 5.415$, with two flavours of staggered
dynamical quarks at $m = 0.025$. This light quark mass corresponds to
a pseudoscalar-to-vector meson mass ratio of $m_{ps}/m_{v} \simeq
0.45$. The lattice volume of these
configurations is $16^{3} \times 32$---with a spacing of $a\sim 0.16$
fm (determined from the charmonium spectrum as described below), this corresponds
to a lattice extending roughly 2.5 fermi in each spatial direction.

We produced an ensemble of quenched configurations with both the
Landau and plaquette tadpole definitions, using the
improved action in Equation \ref{ImpGaugeAction}. We found that, using
Landau and plaquette tadpoles respectively, $\beta = 2.1$ and $\beta = 2.52$ give
almost the same lattice spacing as the unquenched configurations.
These results agree with the spacings given in
Reference \cite{TrottierQuarkonium} at the same values of $\beta$.
We created 100 configurations in each case, with lattice volume $12^{3} \times 24$,
the largest we were able to manage with our computational resources.
Given the small physical size of the heavy mesons, however, the difference in
volume between the quenched and unquenched configurations should not have an
effect on our results.

The lattice spacing was determined for each ensemble using the spin-averaged
$P$--$S$ splitting, for charmonium $E(P-S) = 458$ MeV. This splitting is known
to be quite independent of the heavy quark mass, falling only
slightly to 440 MeV for bottomonium, and so serves as a stable quantity
for determining the physical lattice spacing. We have collected the parameters of our
simulations together in Table \ref{SimulationParameters}.

The \emph{kinetic mass} $M_{k}$ of a boosted state with momentum {\bf p} is defined by
\be
\label{KineticMass}
E({\bf p}) = E(0) + \frac{{\bf p}^{2}}{2M_{k}} + {\cal O}({\bf p}^{4}) \, .
\ee
The bare charm quark mass $M_{0}$ is tuned by requiring that the
kinetic mass of the $^{1}S_{0}$ charmonium state agrees with the
experimental mass of the $\eta_{c}$, $M_{\eta_{c}} = 2.98$ GeV.
We created correlators for a boosted state with ${\bf p} = (\frac{2\pi}{L},0,0)$,
where $L$ is the spatial extent of the lattice. The tuned bare masses, and
their corresponding physical (kinetic) masses for the $^{1}S_{0}$, are shown in Table
\ref{SimulationParameters}.

Meson correlators were calculated for the various charmonium states,
using smeared meson operators with $n_{s} = 8$ and $\epsilon = 1/12$
in Equation (\ref{smearing}) at both
the source and sink. To decrease statistical
uncertainties, we calculated more than one meson correlator for each
gauge field configuration. Meson sources were situated at four different spatial
origins---$(0,0,0)$, $(L/2,L/2,0)$, $(L/2,0,L/2)$ and $(0,L/2,L/2)$---and
starting from two time slices, at $t = 0$ and $t = 12$, for a total of
800 meson correlator measurements for each state.

Statistical correlations will exist between the multiple measurements of the
propagators within each configuration,
however the small size of $Q\bar{Q}$ systems (the $c\bar{c}$ is roughly 0.5 fm
in radius) is some justification for this practice.
The correlations are expected to be small, as noted in other charmonium
studies with similar lattice spacings \cite{TrottierQuarkonium,DaviesCharm}.

Masses for the various $c\bar{c}$ states were found by fitting the
correlators with a single exponential,
\be
G_{M}(t>t_{min}) = c_{M}\, e^{-E_{M}t} \,
\ee
after a minimum time $t_{min}$, allowing for suitable suppression of
excited state contributions.
Energy splittings between correlated states, such as the $S$-state
hyperfine splitting $\Delta E = E(\,^{3}S_{1}) - E(\,^{1}S_{0})$,
can often be extracted more precisely by fitting to
a ratio of their two correlators,
\be
\label{SplittingRatio}
{\cal R}(t) = \frac{G_{B}(t)}{G_{A}(t)}
\rightarrow \frac{c_{B} \, e^{-E_{B}t}}{c_{A} \, e^{-E_{A}t}}
= \frac{c_{B} \, e^{-(E_{A}+\delta E)t}}{c_{A} \, e^{-E_{A}t}}
= c_{{\cal R}} e^{-\delta E t} \, .
\ee
We used ratio fits to extract the $S$-state hyperfine splitting,
and the kinetic mass from the boosted $^{1}S_{0}$ state. Attempts
to extract $P$-state hyperfine splittings in this manner were
unsuccessful, as statistical noise overtook the very small signal
before a reasonable plateau emerged. Single-exponential fits,
however, resolved the three $^{3}P$ levels. We have not employed a
bootstrap analysis for the fit results, which may suggest we have
overestimated the statistical uncertainties.

In the following sections, we present the results for a range of simulations,
incorporating all combinations of quenched and
unquenched gauge configurations, ${\cal O}(v^{4})$ and ${\cal O}(v^{6})$
correction terms, and Landau and plaquette tadpole factors.

\subsection{Quenched results}

\begin{figure}
\begin{center}
\scalebox{0.75}[0.75]{\includegraphics{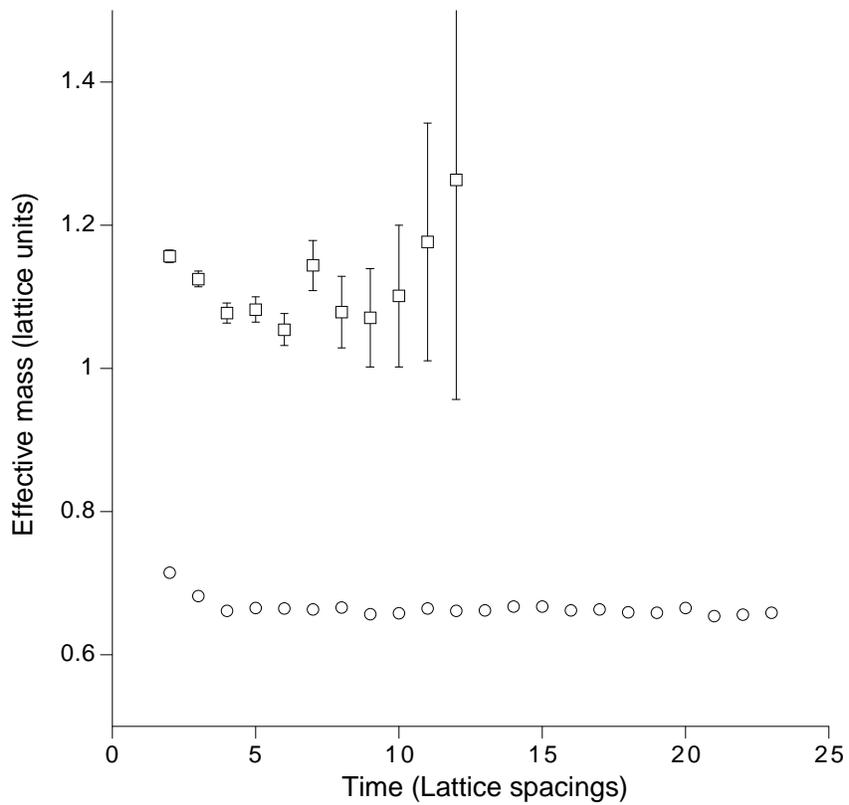}} \caption
{\label{QCharmEff}The effective mass, $- \ln \left (
\frac{G(T+1)}{G(T)} \right )$, of the $^{1}S_{0}$ (circles) and
$^{1}P_{1}$ (squares) in the quenched simulation.}
\end{center}
\end{figure}

\begin{figure}
\begin{center}
\scalebox{0.75}[0.75]{\includegraphics{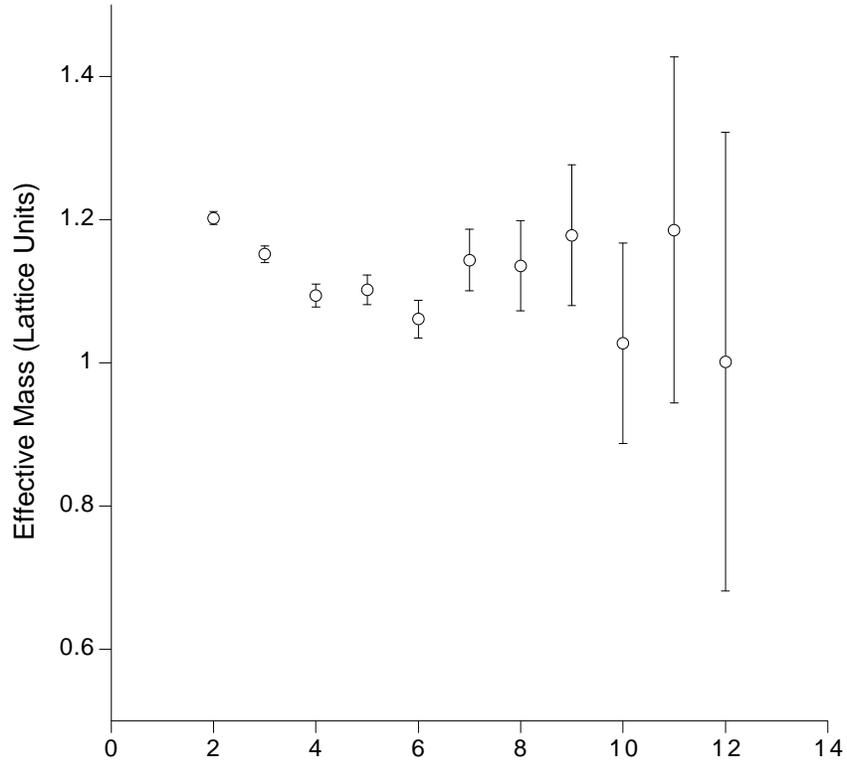}} \caption
{\label{3P0Eff}The effective mass, $- \ln \left (
\frac{G(T+1)}{G(T)} \right )$, of the $^{3}P_{0}$ in the quenched
simulation.}
\end{center}
\end{figure}

An example of the quality of the
correlator data is shown in Figures \ref{QCharmEff} and
\ref{3P0Eff}, plots of the effective masses for the $^{1}S_{0}$,
$^{1}P_{1}$ and $^{3}P_{0}$ from the simulation using the Landau
tadpole factor. The meson propagators were fit with single
exponentials over a range of time intervals $(t_{min}:t_{max})$.
An indication of the convergence of these fits is given in Table
\ref{QuenchedFits}, where the fit results are shown for the ${\cal
O}(v^{6})$ simulations using the plaquette tadpole factor. The
results presented in this table are representative of all of the
charmonium spectra we present here. The two $S$-states had a much
cleaner signal than the four $P$-states, evident in the lower
value for $t_{max}$ used for the $P$-state fits.

\begin{figure}
\begin{center}
    \scalebox{0.75}[0.75]{\includegraphics{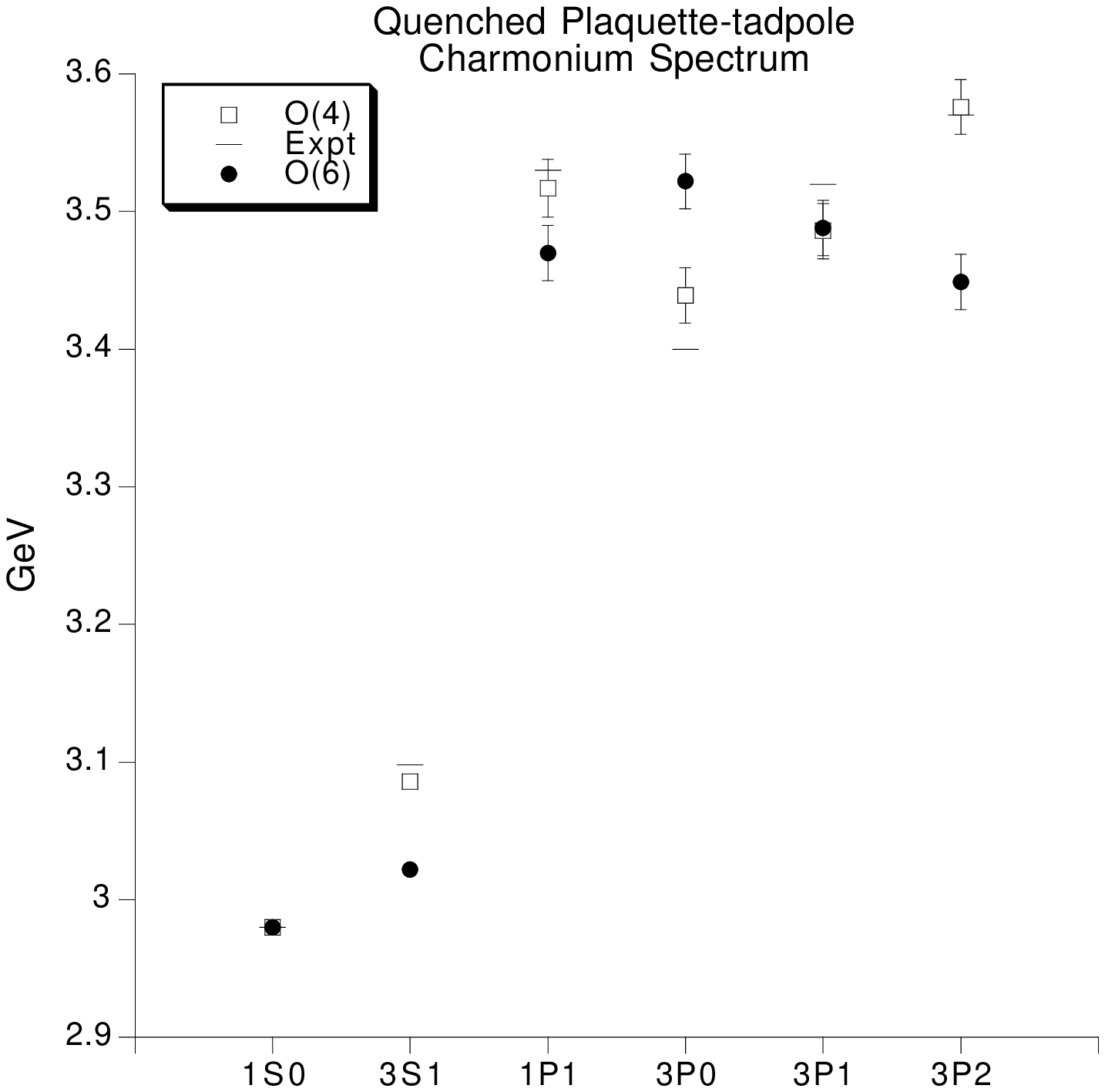}}
\caption{\label{QSpectrum1}Quenched charmonium spectrum using
$u_{0}^{P}$. Squares represent results obtained to ${\cal
O}(v^{4})$, circles represent ${\cal O}(v^{6})$ data. Horizontal
lines indicate experimental values.}
\end{center}
\end{figure}

\begin{figure}
\begin{center}
    \scalebox{0.75}[0.75]{\includegraphics{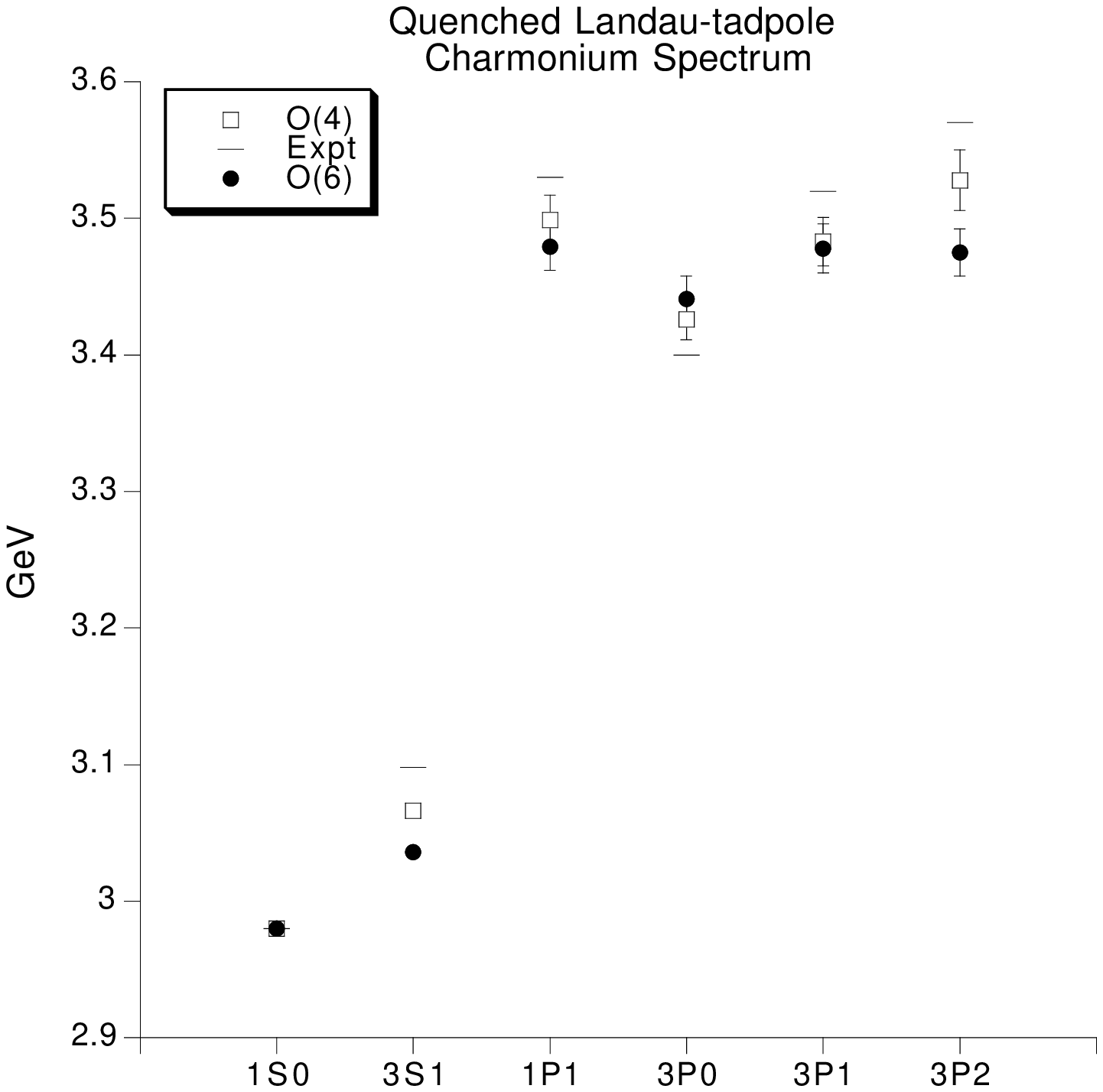}}
\caption{\label{QSpectrum2}Quenched charmonium spectrum using
$u_{0}^{L}$. Squares represent results obtained to ${\cal
O}(v^{4})$, circles represent ${\cal O}(v^{6})$ data. Horizontal
lines indicate experimental values.}
\end{center}
\end{figure}

Table \ref{QuenchedResults} contains
the final results for the quenched charmonium mass fits. We considered the
ground-state for each meson propagator to have properly emerged when three consecutive
$t_{min}:t_{max}$ intervals gave results that agreed within statistical errors;
the meson mass was then taken as the middle of these three values.
The masses are given in both lattice units and physical units,
using the values for $a^{-1}$ in Table \ref{SimulationParameters}
to provide the physical energy scale. The simulated spectra are displayed in
Figures \ref{QSpectrum1} and \ref{QSpectrum2}, shown against the experimental data.

\subsection{Unquenched Results}

\begin{figure}
\begin{center}
    \scalebox{0.75}[0.75]{\includegraphics{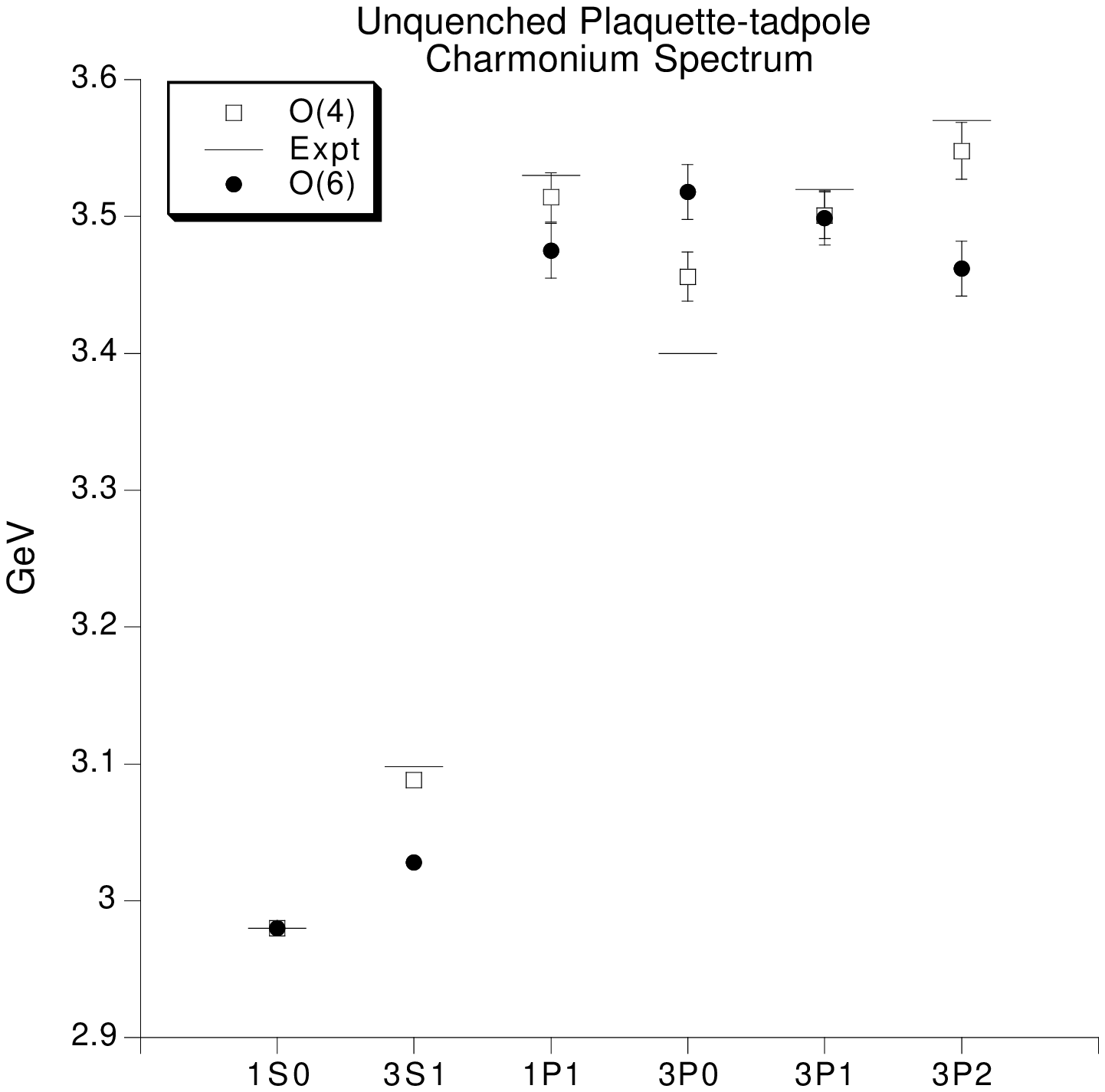}}
\caption{\label{MILCSpectrum1}Unquenched charmonium spectrum using
$u_{0}^{P}$. Squares represent results obtained to ${\cal
O}(v^{4})$, circles represent ${\cal O}(v^{6})$ data. Horizontal
lines indicate experimental values.}
\end{center}
\end{figure}

\begin{figure}
\begin{center}
\scalebox{0.75}[0.75]{\includegraphics{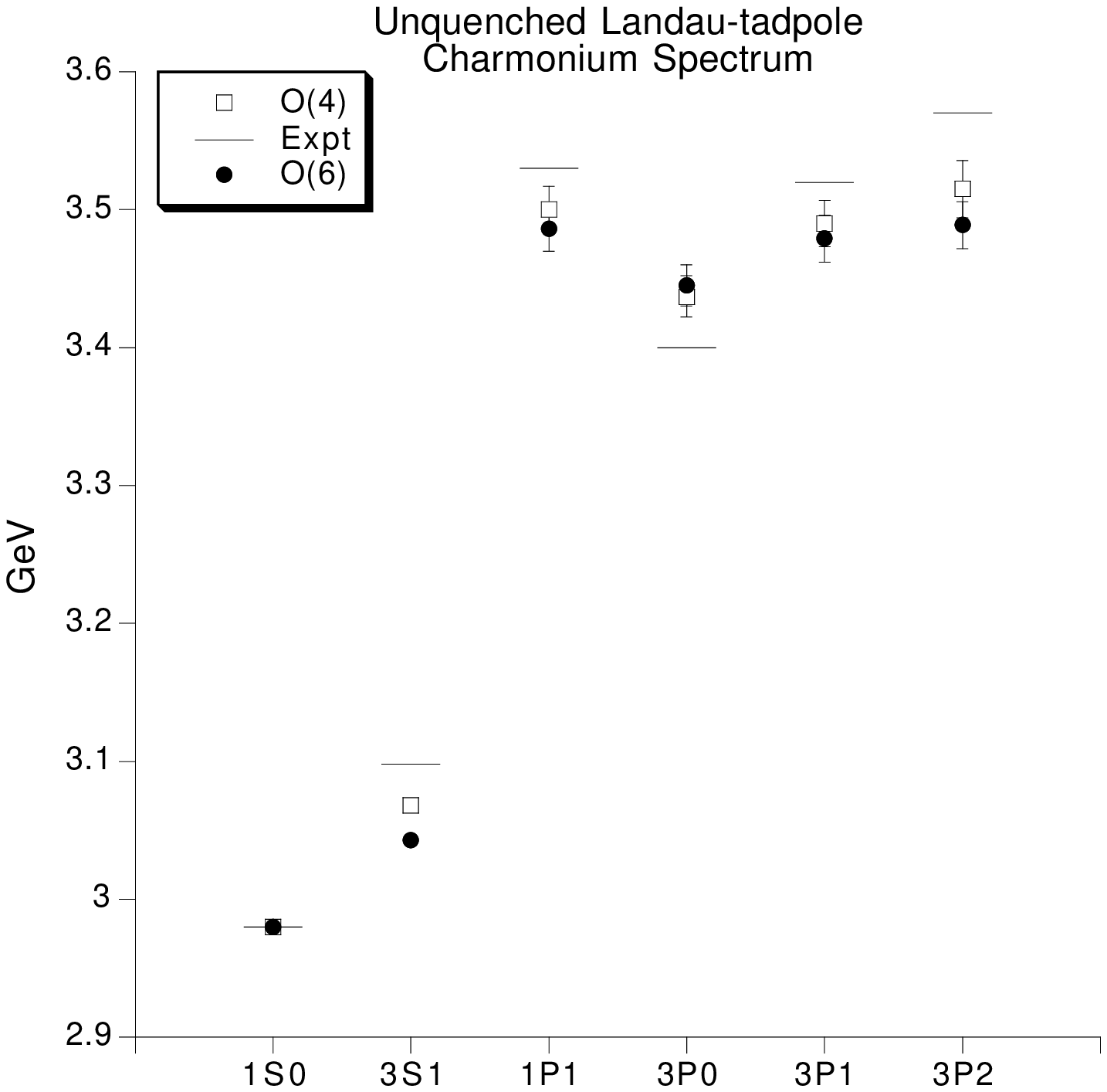}}
    \caption{\label{MILCSpectrum2}Unquenched charmonium spectrum using
$u_{0}^{L}$. Squares represent results obtained to ${\cal
O}(v^{4})$, circles represent ${\cal O}(v^{6})$ data. Horizontal
lines indicate experimental values.}
\end{center}
\end{figure}

Given the similar lattice spacings of the MILC
configurations and our own quenched ensembles, we
have used almost the same parameter set for the unquenched
charmonium simulations---the lower half of Table \ref{SimulationParameters} shows the
specific parameters used. The results of the unquenched simulations are contained in
Table \ref{MILCResults}, with the physical energy scale set
by $a^{-1} = 1.21(2)$ GeV, again from the spin-averaged $P$--$S$ splitting.
The spectra are shown in Figures \ref{MILCSpectrum1} and \ref{MILCSpectrum2}.

\section{Discussion of the Spectra}

\begin{figure}
\begin{center}
\scalebox{0.75}[0.75]{\includegraphics{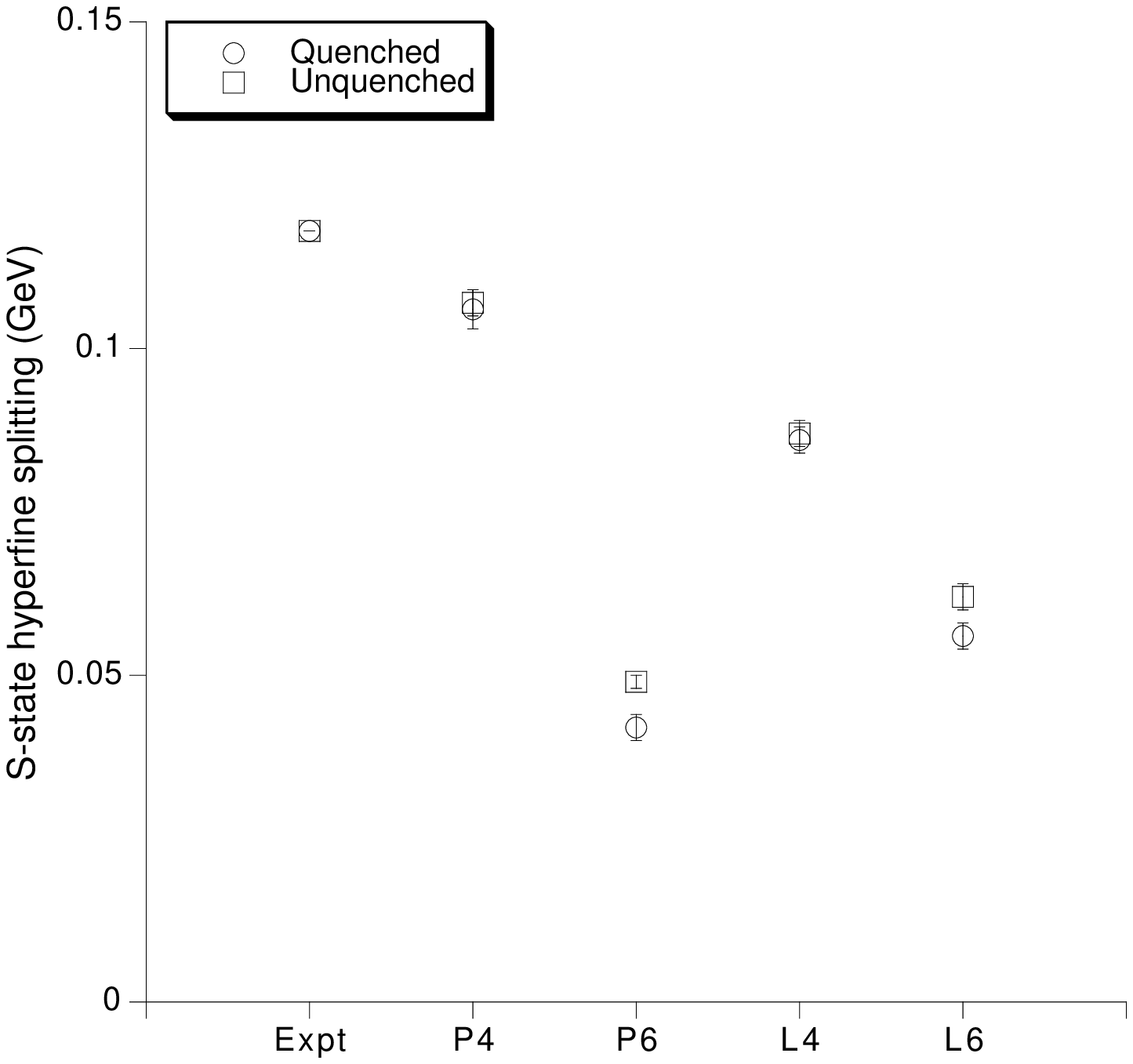}} \caption
{\label{SHyp}Charmonium $S$-state hyperfine splitting. `P4', `P6'
refer to the ${\cal O}(v^{4})$, ${\cal O}(v^{6})$ results obtained
with the plaquette tadpole factor; `L4' and `L6' are the Landau
tadpole results.}
\end{center}
\end{figure}

\begin{figure}
\begin{center}
    \scalebox{0.75}[0.75]{\includegraphics{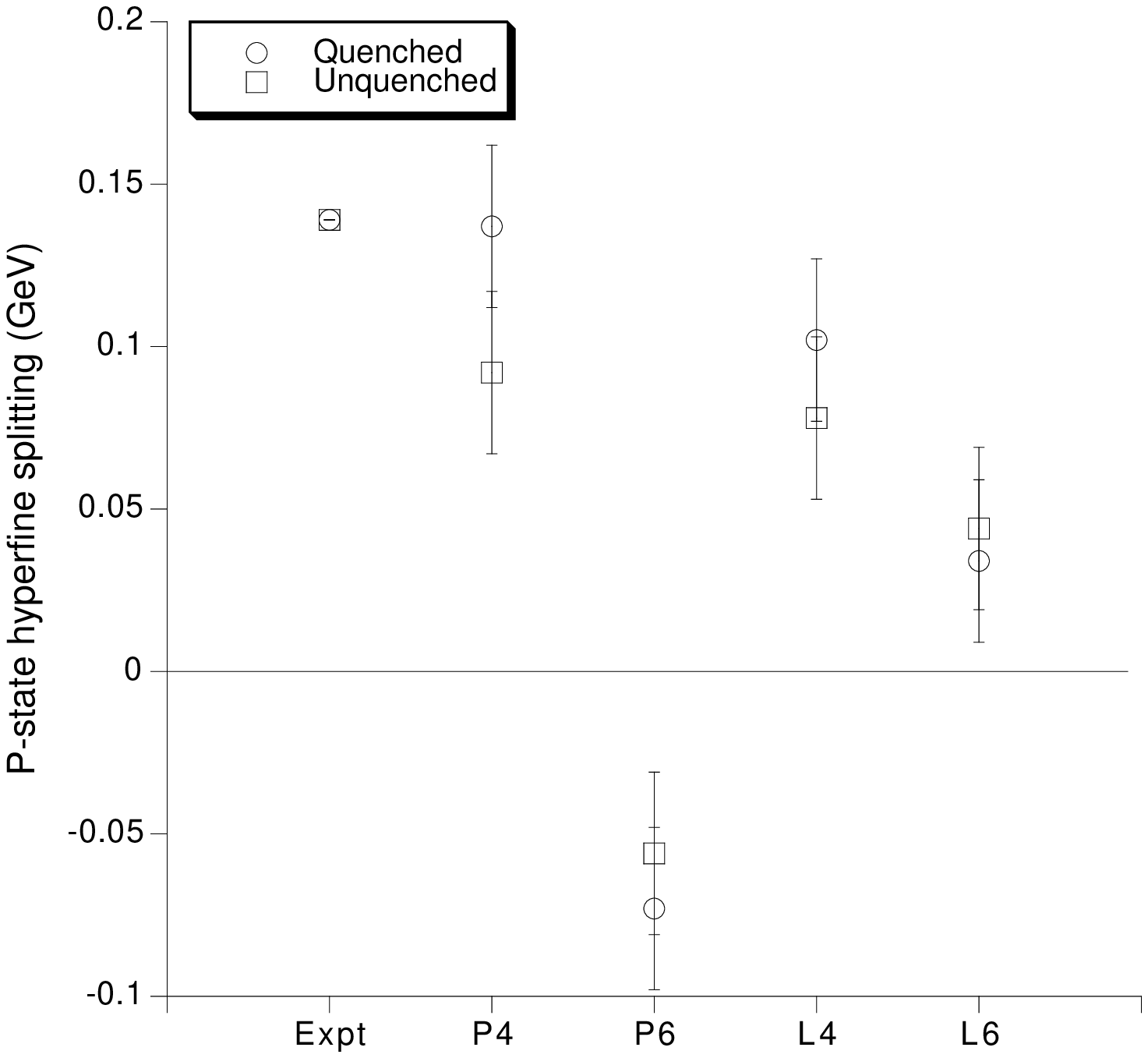}}
\caption{\label{PHyp}Charmonium $P$-state hyperfine splitting.
`P4', `P6' refer to the ${\cal O}(v^{4})$, ${\cal O}(v^{6})$
results obtained with the plaquette tadpole factor; `L4' and `L6'
are the Landau tadpole results.}
\end{center}
\end{figure}

A cursory comparison of the quenched and unquenched results shows
that, while the qualitative structure of the spectrum appears,
precision NRQCD simulations of the charmonium system have a number of
issues yet to be resolved. This is most readily seen in the hyperfine splittings, which are
collected in Figures \ref{SHyp} and \ref{PHyp}, and compared in Table \ref{Hyperfine}.

Consider first the quenched results. The \ov corrections
lead to a disturbingly large decrease in the hyperfine splittings, taking them
further away from the experimental values by as much as $60\%$. The
situation for the plaquette-tadpole simulations is strikingly bad,
where the $^{3}P$ states appear in the wrong order. This
reversal is corrected in the Landau-tadpole simulations, though
the hyperfine splittings are still badly underestimated.

These difficulties are not new---Trottier
\cite{TrottierQuarkonium} first drew attention to the large ${\cal
O}(v^{6})$ corrections to the $S$-state hyperfine splitting in
1996, and noted a possible problem with the $^{3}P$-state
ordering.
Trottier and Shakespeare \cite{Howard} examined the
effects of the different tadpole definitions $u_{0}^{P}$ and
$u_{0}^{L}$ on the $S$-state hyperfine splitting. They performed
${\cal O}(v^{6})$-improved NRQCD simulations using both tadpole
schemes, across a wide range of lattice spacings, and drew a
number of important conclusions; most notably, the ${\cal O}(v^{6})$ hyperfine
corrections with Landau tadpoles were significantly smaller than
the plaquette tadpole results.

We have confirmed a number of these results here, and in
particular clearly resolved the extremely poor $^{3}P$-state
behaviour, most notably when $u_{0}^{P}$ is used. This may simply
be a problem due to the bare charm mass falling below one in these
simulations. However, the $u_{0}^{L}$ simulations lead to a higher
bare $c$-quark mass for a given lattice spacing, and the very low
$P$-state hyperfine splitting even with $aM_{0} > 1$ suggests that these
problems extend beyond the size of the bare mass.

\subsection{Evidence for Quenching Effects?}
\label{DynamicEffects}

The large discrepancies in spin-dependent splittings would be
less worrisome if quenching were seen to have a considerable
effect on the spectrum, as suggested in \cite{charm92}. Sadly,
this does not seem to be the case. There is some evidence
for a difference between the quenched and unquenched simulations in the
\ov $S$-hyperfine data, perhaps as much as ten percent.
However, given the apparent size of other systematic uncertainties, no great
significance can be attached to these differences.

We must address the difference between the quenched and unquenched
gluon actions---the MILC configurations were created with the
Wilson plaquette action, while we have employed the
rectangle-improved action for the quenched lattices. We therefore
anticipate an ${\cal O}(a)$ error entangled with the effects of
the dynamical quarks. Our quenched ${\cal O}(v^4)$ results can be
compared with the results from Reference \cite{DaviesCharm}, where
the plaquette action was used at roughly the same lattice spacing.
We see a $\sim 10$ MeV difference between the $S$-hyperfine
splittings in the two simulations.

We wish to reiterate our goal, however, to see whether the
dynamical quark effects are \emph{large} or \emph{small}. The
$S$-hyperfine splitting, even in relativistic simulations, falls
short of experiment by 40 to 50 MeV. An unquenching effect of this
magnitude would be visible, even taking differences in gluon
action into account. No such effect was observed in these
simulations, and we therefore suggest that quenching effects are
small in this sense.

This conclusion is supported by results in high-precision
$\Upsilon$ simulations \cite{CPPACS00,UnqUps}, where the $P$-state
hyperfine splitting is still somewhat underestimated in unquenched
simulations of this highly-nonrelativistic system, despite the use
of the \ov improved NRQCD action. Very recently, a $10 \%$
sea-quark effect was seen in the hyperfine splittings of the
charmonium and bottomonium system in Reference \cite{CPPACS00},
but differences between the $n_{f} = 0$ and $n_{f} = 2$ $P$-state
splittings were not significant compared with other systematic
uncertainties. Recent results with unquenched lattices in the $B$
meson spectrum have also shown no significant differences between
$n_{f} = 0$ and $n_{f} = 2$ dynamical quark flavours
\cite{unqB99}.

\subsection{Other Systematic Errors}

The preceding results suggest that agreement between lattice
simulations and experiment in quarkonium systems will likely not improve
through the effects of dynamical quarks alone. In the
remainder of this section we explore various other systematic errors
that impact on heavy-quark simulations, as a contrast to the
small quenching effects found above.

\subsubsection{The Choice of the Tadpole Factor}

We have seen, as others have previously, large differences between
results using the Landau tadpole factor $u_{0}^{L}$, and those with the
plaquette definition $u_{0}^{P}$. In our own simulations, the size of the
${\cal O}(v^{6})$ corrections with $u_0^L$ are significantly smaller than the plaquette
tadpole results. This is not surprising: the {\bf E} and {\bf B}
fields are each multiplied by a factor of $u_0^{-4}$ in the tadpole-improved theory.
On our lattices,
\be
\left ( \frac{u_0^P}{u_0^L} \right )^4 = \left \{
\begin{array}{l} 1.24 \,\,\, \mbox{(Quenched)} \\  1.30 \,\,\, \mbox{(MILC)}
\end{array} \right .
\ee
Terms in the NRQCD Hamiltonian linear in {\bf E} or {\bf B} will differ by
as much as $30\%$ between the different tadpole improvement schemes.

As noted earlier, the evidence in favour of Landau tadpoles is
strong. Our simulations offer further support, particularly in the
$^{3}P$-state behaviour, though the more
salient issue here is that tadpole effects are at least as
important as quenching effects in our simulations.

\subsubsection{Radiative Corrections}

We expect some effect on the spectrum
from high-momentum modes that are cut off by the finite lattice spacing.
These high-energy effects may be calculated in perturbative QCD as
${\cal O}(\alpha_{s})$ radiative corrections to the coefficients
of the NRQCD expansion, and there are indications that these may be large
for the charm quark. Lattice perturbation theory calculations of
corrections to $c_{1}$ and $c_{5}$, the `kinetic'
terms in Equation (\ref{NRQCDHamiltonian}), have been completed by Morningstar
\cite{MorningstarPertNRQCD}. The corrections are roughly $10
\% $ or less for the bottom quark, but rise dramatically as the bare
quark mass falls below one (in lattice units). In typical simulations,
the bare charm quark mass sits close to unity, and so these
corrections may become quite significant.

It is possible to find these radiative corrections without
performing long calculations in lattice perturbation theory, by using
Monte Carlo simulations at very high values of $\beta$ \cite{HighBeta}.
Such `non-perturbative' perturbative results have been obtained by Trottier
and Lepage \cite{TrottierHighBeta} for the spin-dependent
$c_{4}$ term in the ${\cal O}(v^{4})$ NRQCD Hamiltonian, Equation
(\ref{NRQCDHamiltonian}). Unfortunately, radiative corrections to the
remaining terms in the NRQCD Hamiltonian have not been calculated to date.

We performed a `toy' simulation to roughly estimate the effects of ${\cal O}(\alpha_{s})$
corrections to all terms in the NRQCD Hamiltonian, replacing the tree level coefficients
$c_{i} = 1$ with $c_{i} = 1 \pm \alpha_{s}$. A rough estimate of $\alpha_{s}$
can be made from the (tadpole-improved) parameters of our simulations,
\be
\alpha_{s}(\pi/a) \simeq \alpha_{lat}^{TI} + {\cal O}(\alpha^{2})
\simeq \frac{g^{2}}{4\pi}
= \frac{6}{4 \pi \beta u_{0}^{4}} \, .
\ee
For our values of $\beta$ and $u_{0}$, this gives $\alpha_{s} \sim
0.15$--$0.2$. For the three terms in the Hamiltonian where
perturbative analysis has been performed, we used the calculated
values \cite{MorningstarPertNRQCD,TrottierHighBeta};
for the remaining terms, we varied the coefficients between 0.8 and 1.2.

Altering the coefficients in this way, we found that the charmonium
$S$- and $P$-hyperfine splittings changed by as much as $10$--$40 \%$, depending
on the sign of the corrections for each individual $c_{i}$.
While this is only a crude estimate, it is clear that the effects
of radiative corrections may be as important as quenching effects for
heavy-quark systems. Accurate determinations of the remaining
${\cal O}(\alpha_{s})$ corrections are sorely needed.

\subsubsection{Improving the Evolution Equation}

The evolution equation we presented in Section \ref{Theory}
for the heavy-quark propagator, Equation (\ref{EvolutionEquation}),
contains better-than-linear
approximations to the exponential $e^{Ht}$ for the terms involving the
zeroth-order Hamiltonian $H_{0}$, but only a linear approximation for the
correction terms $\delta H$. Noting that the high-order corrections
are quite large for charmonium, it is conceivable that this lowest-order
approximation is too severe. A similar conclusion was made by Lewis
and Woloshyn of their NRQCD simulations
of the $D$ meson spectrum \cite{RandyRichard}. The authors were able to remove
some spurious effects due to large vacuum expectation values of one of the
high-order terms in their NRQCD Hamiltonian \cite{RandyRichardNote},
by improving the exponential
approximation for the $\delta H$ terms in the evolution equation.

The coefficients of the \ov terms include high powers
of $M_{0}^{-1}$ and $u_{0}^{-1}$, and it is conceivable that for the
charm quark, with $aM_{0} \sim 1$, the $(1-a\delta H_{v^{6}})$
approximation is poor.
We examined this possibility for the \ov terms, by using an improved form for
the evolution equation that incorporates a `stabilisation' parameter for the
correction terms, with the replacement
\be
(1-a\delta H) \to \left (1-\frac{a\delta H}{s_{\delta}} \right
)^{s_{\delta}} \, .
\ee

We have performed a simulation with this alteration to the
evolution equation, with $s_{\delta} = 4$. Otherwise, all other
parameters were kept the same as the previous Landau-tadpole
quenched simulations. In general, altering the evolution equation
will lead to a change in the bare charm quark mass $M_{0}$. In
this case we found that $M_{0} = 1.15$ once again gave a value of
$3.0(1)$ GeV for the $^{1}S_{0}$ mass.

The the improved evolution equation altered the $S$-hyperfine
splitting significantly, increasing it by roughly $40 \%$ to $70$
MeV. The statistical uncertainties in the $P$-hyperfine splittings
were large, though a similar increase seems likely. These results
suggest the linear approximation $(1- a\delta H)$ typically used
in NRQCD simulations is not sufficiently accurate for the large
corrections encountered at the charm quark mass.

\section{Conclusions}

Lattice NRQCD simulations of heavy-quark systems have evolved
greatly over the last decade. By incorporating high-order
interaction terms to counter relativistic and discretisation
errors, simulations now routinely produce results that agree with
experiment at the $10$--$30 \%$ level. However, stubborn
discrepancies remain in highly-improved simulations, typically
performed in the quenched approximation, or at tree-level in the
${\cal O}(\alpha_{s})$ expansion, or both. To proceed further, all
remaining systematic errors must be addressed.

Past studies strongly suggest that the NRQCD expansion converges
slowly for the charm quark, with the leading and next-to-leading
order corrections apparently oscillating in sign. To ${\cal
O}(v^{6})$, the hyperfine spin-splittings fall short of
experimental values by $50\%$ or more. Without knowing the
magnitude of the next-order corrections in the velocity expansion,
the question of reducing the disparities in the charmonium
spectrum seems academic.

While the NRQCD approach appears to be problematic for the
charmonium system, relativistic lattice formalisms have their
share of difficulties. Simulations of charmonium with a variety of
quark actions---NRQCD, the Fermilab actions, the D234 action---all
underestimate the $S$-hyperfine splitting by at least $40$ MeV
(see \cite{TrottierQuarkonium} for a good summary).

There are sound reasons for estimating the size of dynamical quark
effects in the charmonium system. Some have suggested the
remaining hyperfine discrepancy is due to quenching; estimates of
the effects of dynamical quark loops range as high as $40 \%$
\cite{charm92}. Our results indicate this is unlikely to be the
case---we find at most a $10 \%$ difference between our quenched
and unquenched hyperfine splittings. As the quenching effects are
apparently small for the range of different quark interactions
present in the NRQCD action, we suggest that they will also be
small across other quark actions.

We recognise several shortcomings in our study: we have used
different gluon actions for quenched and unquenched simulations,
we have only examined the effects of unquenching at a single
dynamical quark mass and a single lattice spacing, and we have not
attempted to extrapolate to the physical case of three light sea
quark flavours. The first of these issues was discussed in Section
\ref{DynamicEffects} above. To address the other objections with
further simulations is beyond our present computational resources.
In any case, such efforts are perhaps justified in simulations of
the $b$ quark, where systematic uncertainties are under better
control and quenching effects are probably of comparable size to
discretisation and radiative effects. For the charm system
however, the much larger high-order relativistic errors and the
large tadpole corrections dominate the effects of quenching.

The sensitivity of the NRQCD corrections to the choice of tadpole
factor is well established. This sensitivity should disappear with
a higher-order treatment of the tadpole loops (and other radiative
corrections) in lattice simulations. In practice, such a treatment
is not yet available, and some choice for the tadpole factor is
required. Our results add to the growing list of evidence if
favour of calculating the tadpole correction factor from the mean
link in the Landau gauge, in preference to the plaquette
definition.

The large effects we have encountered due to instabilities in the
evolution equation should also be investigated further. These
instabilities are doubtless amplified in simulations of the charm
quark, where the convergence of the NRQCD expansion is already
questionable. Using an improved evolution equation, as we have
demonstrated, may bring the NRQCD approach into agreement with
other quenched relativistic results for charmonium.

Further, we have shown that ${\cal O}(\alpha_{s})$ radiative
corrections may shift the spin-splittings by as much as $40 \%$.
While this is a crude estimate, the possibility of such sizeable
corrections in comparison with the small quenching effect gives us
pause for consideration. Of particular note are unquenched results
for the $\Upsilon$ spectrum in \cite{UnqUps}, using the ${\cal
O}(v^{6})$ Hamiltonian, which indicate that remaining
discrepancies with experiment are at the ten percent
level---conceivably within the reach of radiative corrections.
Perturbative calculations of the remaining radiative corrections
to the NRQCD coefficients, and those in other actions as well,
will likely be necessary in the near future.

\vspace{0.3cm} We thank Howard Trottier and Randy Lewis for
stimulating discussions. This work has been supported by the
Natural Sciences and Engineering Research Council of Canada.



\newpage



\begin{table}
\begin{center}
\begin{tabular*}{5in}{c @{\extracolsep{\fill}} c c c c c c c} \hline\hline
$\beta$ & $u_{0}^{P}$ & $u_{0}^{L}$ & $a$ (fm) & $a^{-1}$ (GeV) &
$aM_{0}$ & $M_{k}$ & $s$ \\ \hline
 \multicolumn{8}{c}{\emph{Quenched}} \\
2.52 & 0.874 & & 0.168(3) & 1.17(2) & 0.81 & 3.0(1) & 6 \\
2.10 & & 0.829 & 0.181(3) & 1.09(2) & 1.15 & 3.0(1) & 4 \\ \hline
 \multicolumn{8}{c}{\emph{Unquenched}} \\
5.415 & 0.854 & & 0.163(3) & 1.21(2) & 0.82 & 2.9(1) & 6 \\
5.415 & & 0.800 & 0.163(3) & 1.21(2) & 1.15 & 2.9(1) & 4 \\
\hline \hline
\end{tabular*}
\caption{\label{SimulationParameters}
Parameters used in charmonium
simulations. The lattice volume is $12^{3} \times 24$ for the quenched
simulations, and $16^{3} \times 32$ for the unquenched
simulations.
$M_{k}$ is the kinetic mass of the $^{1}S_{0}$ state;
$s$ is the NRQCD stability parameter in Equation
(\ref{NRQCDHamiltonian}).}
\end{center}
\end{table}

\begin{table}
\begin{center}
\begin{tabular*}{5in}{c @{\extracolsep{\fill}} c c c} \hline \hline
$t_{min}:t_{max}$ & $^{1}S_{0}$ & $^{3}S_{1}$ & $^{3}S_{1}$ -
$^{1}S_{0}$ \\ \hline
2:24 & 0.6646(4) & 0.7000(5) & --- \\
3:24 & 0.6630(5) & 0.6984(5) & 0.03659(9) \\
4:24 & \emph{0.6625(5)} & \emph{0.6979(6)} & 0.03661(11) \\
5:24 & 0.6624(6) & 0.6977(7) & \emph{0.03645(13)} \\
6:24 & 0.6623(7) & 0.6976(7) & 0.0364(2) \\
7:24 & 0.6623(7) & 0.6976(8) & 0.0364(2) \\ \hline \hline
\end{tabular*}
\begin{tabular*}{5in}{c @{\extracolsep{\fill}} c c c c}
$t_{min}:t_{max}$ & $^{1}P_{1}$ & $^{3}P_{0}$ & $^{3}P_{1}$ &
$^{3}P_{2}$ \\ \hline
2:14 & 1.109(4) & 1.159(4) & 1.138(5) & 1.082(4) \\
3:14 & 1.093(5) & 1.127(7) & 1.113(6) & 1.072(5) \\
4:14 & \emph{1.085(7)} & \emph{1.122(10)} & \emph{1.102(9)} & \emph{1.065(6)} \\
5:14 & 1.087(10) & 1.139(15) & 1.107(12) & 1.067(9) \\
6:14 & 1.091(13) & 1.14(2) & 1.114(18) & 1.067(12) \\
\hline \hline
\end{tabular*}
\caption{\label{QuenchedFits}
Examples of fits to quenched charmonium propagators. The
fit results are shown for the plaquette-tadpole simulation, for
various sets of $(t_{min}:t_{max})$. Single exponential fits are used
for individual masses, and a ratio fit is used to extract the
$S$-state hyperfine splitting. Italicised entries indicate the final
results.}
\end{center}
\end{table}

\begin{table}
\begin{center}
\begin{tabular*}{5in}{c @{\extracolsep{\fill}} c c c c} \hline \hline
State  & \multicolumn{2}{c}{$u_{0}^{P}$} & \multicolumn{2}{c}{$u_{0}^{L}$} \\
  & ${\cal O}(v^{4})$ & ${\cal O}(v^{6})$
  & ${\cal O}(v^{4})$ & ${\cal O}(v^{6})$ \\ \hline
$^{1}S_{0}$ & 0.5733(5) & 0.6625(5) & 0.1708(4) & 0.2297(4)  \\
$^{3}S_{1}$ & 0.6635(8) & 0.6979(6) & 0.2466(6) & 0.2802(5)  \\
$^{1}P_{1}$ & 1.034(8) & 1.085(7) & 0.643(7) & 0.696(7)  \\
$^{3}P_{0}$ & 0.966(7) & 1.122(10) & 0.576(6) & 0.661(7)  \\
$^{3}P_{1}$ & 1.006(8) & 1.102(9) & 0.628(7) & 0.695(8)  \\
$^{3}P_{2}$ & 1.088(8) & 1.065(6) & 0.669(9) & 0.692(7)  \\
$^{3}S_{1} -$$^{1}S_{0}$ & 0.0910(3) & 0.0365(1) & 0.0778(3) & 0.0521(2) \\
\end{tabular*}
\begin{tabular*}{5in}{c @{\extracolsep{\fill}} c c c c c}
\hline \hline
State  & \multicolumn{2}{c}{$u_{0}^{P}$} & \multicolumn{2}{c}{$u_{0}^{L}$}
& Expt \\
  & ${\cal O}(v^{4})$ & ${\cal O}(v^{6})$
  & ${\cal O}(v^{4})$ & ${\cal O}(v^{6})$ & \\ \hline
$^{3}S_{1}$ & 3.086(2) & 3.022(2) & 3.066(2) & 3.036(1) & 3.097 \\
$^{1}P_{1}$ & 3.517(17) & 3.470(17) & 3.499(18) & 3.479(17) & 3.526 \\
$^{3}P_{0}$ & 3.439(16) & 3.522(20) & 3.426(15) & 3.441(17) & 3.417 \\
$^{3}P_{1}$ & 3.486(17) & 3.488(17) & 3.483(18) & 3.478(18) & 3.511 \\
$^{3}P_{2}$ & 3.576(17) & 3.449(16) & 3.528(22) & 3.475(17) & 3.556 \\
$^{3}S_{1} -$$^{1}S_{0}$ & 0.106(3) & 0.042(2) & 0.086(2) &
0.056(2) & 0.118 \\
\hline \hline
\end{tabular*}
\caption
{\label{QuenchedResults}
Quenched charmonium masses in lattice units (top) and GeV
(bottom). The scale is set
by $a^{-1}$ in Table \ref{SimulationParameters}.}
\end{center}
\end{table}

\begin{table}
\begin{center}
\begin{tabular*}{5in}{c @{\extracolsep{\fill}} c c c c} \hline \hline
State  & \multicolumn{2}{c}{$u_{0}^{P}$} & \multicolumn{2}{c}{$u_{0}^{L}$} \\
  & ${\cal O}(v^{4})$ & ${\cal O}(v^{6})$
  & ${\cal O}(v^{4})$ & ${\cal O}(v^{6})$ \\ \hline
$^{1}S_{0}$ & 0.5501(3) & 0.6279(3) & 0.0581(3) & 0.1155(3)  \\
$^{3}S_{1}$ & 0.6363(5) & 0.6668(4) & 0.1278(4) & 0.1658(4)  \\
$^{1}P_{1}$ & 0.988(7) & 1.030(10) & 0.485(7) & 0.537(6)  \\
$^{3}P_{0}$ & 0.937(6) & 1.020(8) & 0.433(6) & 0.503(6)  \\
$^{3}P_{1}$ & 0.977(7) & 1.050(10) & 0.476(7) & 0.531(7)  \\
$^{3}P_{2}$ & 1.016(9) & 1.065(5) & 0.497(9) & 0.540(7)  \\
$^{3}S_{1} -$$^{1}S_{0}$ & 0.0884(3) & 0.0365(2) & 0.0719(2) &
0.0522(1) \\
\end{tabular*}
\begin{tabular*}{5in}{c @{\extracolsep{\fill}} c c c c c}
\hline \hline
State  & \multicolumn{2}{c}{$u_{0}^{P}$} & \multicolumn{2}{c}{$u_{0}^{L}$} & Expt \\
  & ${\cal O}(v^{4})$ & ${\cal O}(v^{6})$
  & ${\cal O}(v^{4})$ & ${\cal O}(v^{6})$ \\ \hline
$^{3}S_{1}$ & 3.087(2) & 3.028(2) & 3.068(2) & 3.043(1) & 3.097 \\
$^{1}P_{1}$ & 3.514(17) & 3.475(17) & 3.500(17) & 3.486(16) & 3.526 \\
$^{3}P_{0}$ & 3.456(15) & 3.518(20) & 3.437(15) & 3.445(15) & 3.417 \\
$^{3}P_{1}$ & 3.501(17) & 3.499(17) & 3.490(17) & 3.479(17) & 3.511 \\
$^{3}P_{2}$ & 3.548(21) & 3.462(16) & 3.515(21) & 3.489(17) & 3.556 \\
$^{3}S_{1} -$$^{1}S_{0}$ & 0.107(2) & 0.049(1) & 0.087(2) &
0.062(1) & 0.118 \\
\hline \hline
\end{tabular*}
\caption
{\label{MILCResults}
Unquenched charmonium masses in lattice units (top) and GeV
(bottom). The scale is set by $a^{-1}$ in Table \ref{SimulationParameters}.}
\end{center}
\end{table}

\begin{table}[t]
\begin{center}
\begin{tabular*}{5in}{c @{\extracolsep{\fill}}c c c c c}\hline \hline
&  &  & $S$-state hyperfine & $P$-state hyperfine \\
&  &  & $^{3}S_{1}-$$^{1}S_{0}$ & $^{3}P_{2}-$$^{3}P_{0}$ \\ \hline
Experiment & & & 0.118 & 0.139 \\ \hline
${\cal O}(v^{4})$ & Quenched & $u_{0}^{P}$ & 0.106(2) & 0.14(2) \\
 & & $u_{0}^{L}$ & 0.086(2) & 0.10(2) \\
 & Unquenched & $u_{0}^{P}$ & 0.108(2) & 0.10(2)\\
 & & $u_{0}^{L}$ & 0.087(2) & 0.08(2)\\ \hline
${\cal O}(v^{6})$ & Quenched & $u_{0}^{P}$ & 0.042(1) & -0.07(2)\\
 & & $u_{0}^{L}$ & 0.056(1) & 0.033(15)\\
 & Unquenched & $u_{0}^{P}$ & 0.049(1) & -0.023(17)\\
 & & $u_{0}^{L}$ & 0.062(1) & 0.044(15)\\
\hline \hline
\end{tabular*}
\caption[Quenched and unquenched $S$-state and $P$-state hyperfine splittings]
{\label{Hyperfine}
Quenched and unquenched $S$-state and $P$-state hyperfine splittings
for both $u_{0}^{P}$ and $u_{0}^{L}$ simulations.}
\end{center}
\end{table}

\clearpage



\begin{thebibliography}{99}
\bibitem{RandyRichardNote} The term considered in \cite{RandyRichard}
appears at ${\cal O}(v^{8})$ in the NRQCD expansion, so we have not
included it in our simulations.
\end{thebibliography}
\end{document}